# Oscillations in networks of networks stem from adaptive nodes with memory

Amir Goldental[1,†], Herut Uzan[1,†], Shira Sardi[1] and Ido Kanter[1,2,*]

[1]Department of Physics, Bar-Ilan University, Ramat-Gan 52900, Israel
[2]Gonda Interdisciplinary Brain Research Center and the Goodman Faculty of Life Sciences, Bar-Ilan University, Ramat-Gan 52900, Israel
†These authors contributed equally to this work.
*Corresponding author: ido.kanter@biu.ac.il

**We present an analytical framework that allows the quantitative study of statistical dynamic properties of networks with adaptive nodes that have memory and is used to examine the emergence of oscillations in networks with response failures. The frequency of the oscillations was quantitatively found to increase with the excitability of the nodes and with the average degree of the network and to decrease with delays between nodes. For networks of networks, diverse cluster oscillation modes were found as a function of the topology. Analytical results are in agreement with large-scale simulations and open the horizon for understanding network dynamics composed of finite memory nodes as well as their different phases of activity.**

Networks and networks of networks are important concepts in many fields of physics[1-3]. The quantitative study of such complex systems presents a better understanding of the physical process in a verity of fields[4], e.g. lasers[5], power grids[6] and even economics[7]. Common questions address the structural properties of networks or the dynamics within the networks, given the statistics of the interactions, links. However, the complexity of the analytical approach increases immensely when the activity of the nodes is not trivial. Here we examine a model where nodes have memory and their responses are dependent on the history of the activity of connected nodes.

We develop an analytical framework, which allows the examination of the statistical properties of such networks and their dynamics, based on stochastic equations in the mean-field limit. We test our analytical framework on a model where nodes have activity-dependent response probability. Although this model is based on experimental results on neural networks[8-11], their dynamics resemble many other physical systems[12]. Results are quantitatively found to be in a good agreement with large-scale simulations and support that the examined model leads to collective network oscillations. Generalization to more structured topologies, consisting of networks of networks[13,14], is presented, indicating additional diverse intra-cluster modes of collective oscillations which are determined by the topology of the complex network[15].

## Results

**The model**. In the manuscript we examine a network of firing neurons, where neurons sum up the inputs they receive and respond stochastically if a threshold is crossed. The probability for a neuronal response is a function of the previous activity of its neighbors and the several previous threshold crossings of the neuron. In the general case, the dynamics are in continues time, however, if all the delays between connected neurons are equal, the situation can be simplified to an iterative map.

Excitatory neural networks can be modeled using leaky integrate and fire neurons[16,17]. The dynamics of each neuron within the network is then described by a voltage equation

$$\frac{dV_i}{dt} = -\frac{V_i - V_{st}}{\tau} + \sum_{j=1}^{N} w_{ji} \sum_{n} \delta(t - t_j(n) - d_{ji}) \quad (1)$$

where *i* indicates the neuron indexing, $w_{ji}$ and $d_{ji}$ are the connection's strength and delay from neuron *j* to *i*, respectively, *N* stands for the number of neurons and the values used are $\tau$=20 ms for the membrane time constant and $V_{st}$=-70 mV for the stable membrane (resting) potential. The summation over $t_j(n)$ sums all the past firing times of neuron *j* and spontaneous stochastic synaptic currents, i.e. external noise. Notice that when several spikes arrive simultaneously, they sum up to a single input pulse with a larger amplitude. A neuronal threshold is typically defined at $V_{th}$=-54 mV and a threshold crossing results in an evoked spike followed by a refractory period of a few milliseconds, e.g. *2 ms*. During this period no evoked spikes are possible and the voltage is set to *-75 mV*.

A persistent challenge in neuroscience is that the neuron's membrane voltages change continuously most of the time, but occasionally spikes sharply. Therefore, certain aspects of the dynamics are best described by a discrete-time approach while others require continuous dynamics. When writing a continues-time equation, one needs to take into account that the focal neuron, i, will receive sharp inputs from other neurons, approximate by delta spikes. Since the spikes fired by neuron *j* arrive at a neighbor neuron *i* with some time delay $d_{ji}$ we have to take into account all previously fired spikes, such that the input received by *i* from *j* can be expressed as the sum $\sum_n \delta(t-t_j(n)-d_{ji})$, last term in eq. (1).

In the case where the connections between neurons are supra-threshold, $V_{th}-V_{st}<w_{ji}$, the dynamical evolutions of Eq. (1) is simplified and the only relevant information is whether a neuron fire or not. The case of a network composed of binary neurons is easier to analyze and is at the center of this study, where the extension for sub-threshold connections is discussed later.

The standard version of the presented model is modified, according to recent experimental observations[8,9], by introducing probability for the intrinsic spike generation[8,9]. Experiments report that each neuron is characterized by a critical stimulation frequency, $f_c$, ranging from sub-Hertz to about 30. When a neuron is repeatedly stimulated (supra-threshold), its average firing frequency is saturated at $f_c$. In the case of regular stimulations, with inter-stimulation intervals equal to $\Delta t$, the probability for a spike given a stimulation is

$$p^{sp}(\Delta t) = \min(\Delta t \cdot f_c, 1) \qquad (2)$$

indicating that the response probability is $\Delta t \cdot f_c$, or unity in case $\Delta t \cdot f_c > 1$. Note that the response failures were found to be stochastic and statistically independent.

In the general case of irregular (supra-threshold) stimulations, with non-identical inter-stimulation intervals, experimental results reveal that the response probability depends on the last several stimulations with an exponentially decaying memory[8,9]. Quantitatively, the neuronal response probability for the $n^{th}$ voltage threshold crossing of a neuron is given by

$$p_\alpha^{sp}(\Delta t_2, \Delta t_3, \ldots, \Delta t_n) = p^{sp}((1-\alpha)\sum_{m \leq n} \Delta t_m \alpha^{n-m}) \qquad (3)$$

where *0≤α<1* measures the neuronal memory, $\Delta t_m$ is the time-lag between the *(m-1)$^{th}$* and *m$^{th}$* voltage threshold crossings, $p^{sp}$ is the neuronal response probability for given weighted inter-stimulation-intervals, Eq. (2), and *(1-α)* stands for the normalization factor of the weights, e.g. if all $\Delta t_m = \Delta t$, the *weighted average* is equal to $\Delta t$. In the case of a response failure, the voltage is set to *-65 mV*, response failures are also counted as voltage threshold crossings.

Neural networks are commonly assumed to be composed of many weak connections and sparse strong connections, where several simultaneous stimulations lead to threshold crossings[18,19]. Here we neglect the weak connections and assume solely strong connections ($w_{ji}$), such that each stimulation leads to a threshold crossing.

The parameters of the examined systems, if not stated otherwise, are $N=2000$, $d=10$ ms, $f_c=10$ Hz, $\alpha=0$ and the probability for a neuron to have $k$ random incoming (presynaptic) connections is $C_k=<k>^k \cdot \exp(-<k>)/k!$ as in a sparse ER random graph[3], where $<k>=3$ stands for the average connectivity. The reported qualitative results were found to be insensitive to these parameters and are independent of initial conditions. In addition, stochastic uncorrelated stimulations, imitating spontaneous stochastic synaptic currents, were injected into the system with an average rate of $f_{ext}=0.1$ Hz per neuron, which also excludes the death possibility of the activity of the entire finite network. An example of such a simulation is presented in Fig. 1a.

**Analytical approach.** Using the following mean-field approach, statistical features of the dynamics, e.g. the dominating oscillation frequency[20] (Fig. 1b), are well estimated.

The analytical study is further simplified where all delays are equal to $d$, hence, the time domain is discretized to steps of multiples of $d$. We look on the statistical properties of the activity, such as the probability of a neuron to be stimulated and the probability of a neuron to fire during the time steps. Our analytical framework divides neurons into groups, where each group consists of neurons with a giving incoming connectivity. For each group we define a conditional probability, i.e. the probability for a neuron in that group to be stimulated at a given time step and then to produce an evoked spike. The mean-field approximation of all these conditional probabilities enables the estimation the time evolution of macroscopic quantities such as the average firing rate.

We start from a simplified case without long-term nodal memory ($\alpha<<1$). The quantization of the time domain to multiples of $d$ (time=$i \cdot d$), enables to describe the fraction of firing neurons, $R(i)$, at each time step as

$$R(i) = \sum_{k=0}^{k_{max}} C_k \cdot p_k^{st}(i) \cdot \chi_k(i) \qquad (4)$$

where $k_{max}$ is the maximal possible number of incoming connections per node, $C_k$ is the probability for a neuron of having $k$ incoming connections, $p^{st}_k(i)$ stands for the fraction of stimulated neurons at time step $i$ out of all neurons that have $k$ incoming connections and the susceptibility $\chi_k(i)$ stands for their probability for evoked spike.

Eq. (4) can be interpreted as a simple weighted summation of independent probabilities. Specifically, the probability for a neuron to fire at time step $i$, $R(i)$, is the weighted summation ($C_k$ is the weight function) over the probabilities ($p^{st}_k(i) \cdot \chi_k(i)$) of a neuron to fire at time step $i$, given it has $k$ incoming connections, where the summation is over $k$. The term $p^{st}_k(i) \cdot \chi_k(i)$ is simply the [probability for a stimulation at time step $i$, given the neuron has $k$ incoming connection] multiplied by [the probability for a spike, given a stimulation at time step $i$ to a neuron with $k$ incoming connections].

The fraction $p^{st}_k(i)$ can be estimated as

$$p^{st}_k(i) = 1 - [1 - R(i-1)]^k \cdot (1 - f_{ext} \cdot d) + \xi^{st}_k(i) \qquad (5)$$

where $[1-R(i-1)]^k$ is the probability that no presynaptic neuron evoked a spike at the previous time step, $(1-f_{ext} \cdot d)$ is the probability that a neuron was not stimulated since the last step by synaptic noise and the last term stands for random fluctuations in finite networks which scale with $N^{-0.5}$ (Gaussian random variable with a zero mean and a variance equals to $<p^{st}_k(i)> \cdot (1-<p^{st}_k(i)>)/(C_k \cdot N)$, where $<\cdot>$ averages over the noise term).

Throughout the manuscript, we study the case of homogeneous external stimulations, i.e. $f_{ext}$ is time-independent and the probability of external stimulation is constant in each time step, a Poisson process. As a result, the probability of a neuron to be stimulated in a time step of duration $d$ is $1-\exp(-f_{ext} \cdot d)$, which for small values is well approximated by $f_{ext} \cdot d$ (around *0.001*). However, for non-homogeneous external stimulations, e.g. the external input depends on time, the equations can be generalized to include t time-dependent external stimulation frequency, $f_{ext}(i)$.

The susceptibility, $\chi_k(i)$, is estimated by classifying the nodes into subgroups according to the number of time steps passed since their last stimulations, e.g a group of neurons that were stimulated at step *i-3* but not at *i-2* nor *i-1*. We denote by $h_k(i,m)$ the probability that $m$ time steps passed from the last stimulation for a neuron at time $i$ that has $k$ incoming connections. This quantity, $h_k(i,m)$, can be approximated as a nonhomogeneous Poisson process, i.e. the probability to be stimulated at time step *i-m* multiplied by the probability not to be stimulated at all time steps between *i-m* and *i*:

$$h_k(i,m) = p^{st}_k(i-m) \prod_{n=1}^{m-1} [1 - p^{st}_k(i-n)] + \xi^h_k(i,m) \qquad (6)$$

where the last term stands for random fluctuations in finite networks which scale with $N^{-0.5}$ (Gaussian random variable with a variance of $<h_k(i,m)> \cdot (1-<h_k(i,m)>)/(p^{st}_k(i) \cdot C_k \cdot N)$ and a zero mean). All neurons belonging to $h_k(i,m)$ have a response probability of $p^{sp}(m \cdot d)$, Eq. (2), thus, the susceptibility is:

$$\chi_k(i) = \sum_{m=1}^{\infty} p^{sp}(m \cdot d) \cdot h_k(i,m) \qquad (7)$$

Taking into account that $h_k(i,m)$ is normalized for each $i$ (Eqs. S1 and S2), and that $p^{sp}(m \cdot d) = m \cdot d \cdot f_c$ for all $m < (f_c \cdot d)^{-1}$ and $1$ for $m > (f_c \cdot d)^{-1}$, Eq. (7) can be written as a finite sum:

$$\chi_k(i) = 1 - \sum_{m=1}^{(f_c \cdot d)^{-1}} (1 - m \cdot d \cdot f_c) \cdot h_k(i,m) \qquad (8)$$

In case any of the probabilities stochastically deviates from [0,1], it truncates to the boundaries.

Note that the noisy terms of Eqs. (5) and (6) result from a random selection process, the summation of many Bernoulli trials which leads to a Gaussian random variable, according to the central limit theorem. Specifically, in Eq. (5) the stochastic term is a result of Bernoulli trials where the "success" is a stimulation to a neuron. The stochastic term in Eq. (6) is similar, however, there is a constraint that the sum of $h_k(i,m)=1$, when summing over m from 1 to infinity, thus some non-intuitive correlation appears within the terms. We solved this correlation by renormalizing the terms (Eqs. S1 and S2, supplementary). Nevertheless, a more sophisticated and complicated approach is possible (not shown), where the noise terms are calculated in a recursive way, creating dependencies. Preliminary results indicate that the complicated approach does not lead to a noticeable advantage over the simplified one.

**Comparison of analytical and simulation results.** The fraction of firing neurons, $R(i)$, was found by solving numerically Eqs. (4)-(8) for various networks. A comparison between this analytical approach and simulations is demonstrated in Fig. 1. Results indicate a fair agreement in the power spectrum of the activity with a maxima around 8.3 Hz, which is defined as the oscillation frequency. The dependency of the oscillation frequency on the parameters defining the network is analyzed in Fig. 2. Both analytical and simulation results clearly indicate that the oscillation frequency increases with $f_c$ and with the average incoming connectivity $<k>$ and decreases with the delay, $d$ (Fig. 2a-c).

The case of nodes with a longer memory, i.e. $\alpha$ is not small, as was found experimentally[8], can be also examined using a similar analytical framework with the following generalized Eq. (8):

$$\chi_k(i) = 1 - \sum_{m=1}^{(f_c \cdot d)^{-1}} \left[1 - (1-\alpha) \cdot m \cdot d \cdot f_c - \alpha \cdot \frac{d \cdot f_c}{k \cdot \langle R \rangle_i}\right] \cdot h_k(m) \qquad (9)$$

where $<R>_i$ is the average of $R(i)$ over $i$ and is estimated numerically (Eqs. S3 and S4) as the steady state solution of Eqs. (4)-(8) while setting $R(i)=<R>_i$ for all $i$ and solving without the noise terms. The effect of $\alpha$ on the oscillation frequency is illustrated in Fig. 2d, indicating that the oscillation frequency decreases with $\alpha$.

The reported trends are controlled by the neuronal response failures and the topology of the network. Increasing $<k>$ or decreasing $d$ leads to a faster formation of the oscillations and distribution of activity through the network, hence allowing higher frequency oscillations. Similarly, increasing the saturated firing frequency of the neurons, $f_c$, increases the rate of the information propagation through the network, allowing faster oscillations. On the other hand, increasing $\alpha$ reduces the sensitivity of the nodes to changes in the network activity, challenging the synchronization between nodes and reducing the oscillation frequency.

The comparison between the oscillation frequencies measured in simulations of finite $N$ and mean-field is presented in Figs. 2e-f. Results clearly indicate that at N=2000 the difference is negligible. In addition the finite size scaling, Fig. 2f, indicates a convergence of the error, which results from the finite size, as $N^{-0.5}$, which is probably attributed to the variance of the noise (Eqs. (5)-(6)).

**Network of networks.** The generalization of the presented analytical framework to the case of $n$ connected sub-networks, e.g. Fig. 3, is straightforward. In such a case, the activity of each sub-network is characterized by $R_g(i)$, $g=1,2,…,n$ and the stimulation probability $p^{st}{}_k(i)$ is then changed according to the connectivity among sub-networks, i.e. the probability for a neuron to be stimulated, given it has $k_1, k_2, k_3,..., k_n$ connections from sub-network $1, 2, 3,…, n$, respectively, is

$$p^{st}_{k_1,k_2,…,k_n}(i) = 1 - (1 - f_{ext} \cdot d) \cdot \prod_{g=1}^{n}[1 - R_g(i - d)]^{k_g} + \xi^{st}_{k_1,k_2,…,k_n}(i) \qquad (10)$$

The full set of the modified Eqs. (4)-(8) are presented in Eqs. S5.

An example of the activity of a network of networks, with the topology as in Fig. 3, is presented in Fig. 4. All neurons have 1 incoming connection from their own sub-network with delay $d=10$ ms and 2 incoming connections from the connected sub-networks with delay $D=2d$. This *network of networks* configuration produces cluster zero-lag synchronization[15] which coexists with new oscillation modes within each sub-network.

The activity of each of the four sub-networks is exemplified in Fig. 4a, color coded according to Fig. 3. A power spectrum of the activity of the top-left (orange) network is presented (Fig. 4b), indicating new modes of oscillations among the sub-networks, which coexists with the low-frequency self-oscillations within the dynamics of each sub-network (Fig 2). The emergence of the additional modes of oscillations is attributed to the structure of the entire network, creating two inner-loops among sub-networks, *2D (40 ms, 25 Hz)* and *4D (80 ms, 12.5 Hz)*. These local loops create two dominating non-local zero-lag synchronized clusters[15], blue with green and orange with red. It represents a non-local mechanism for zero-lag synchronization, since the orange and the red sub-networks, for instance, are not directly connected. A direct evidence for the zero-lag synchronization is exemplified by the cross-correlation between the rate of the orange sub-network and the red sub-network (Fig. 4c) emerging simultaneously with a leader-laggard synchronization between the connected, green and blue, sub-networks. Note that the activity of the networks is affected by the intrinsic noise and the driving force from connected networks, which lead to a more noisy and structured activity than of a single network (Fig. 1), and is within the line of reported experimental results[21,22]. The dissipation timescale of emerging fluctuations approximately depends on the deterministic part of Eqs. (1)-(4).

In the presented case, the connectivity between sub-networks is higher than the connectivity within each sub-network, however, the opposite case where the intra-cluster connectivity is above the average is also very interesting[23]. This limit can be studied by using Eqs. S5 (supplementary). Specifically, when the intra-cluster connectivity is above the average connectivity, preliminary results indicate that qualitatively different dynamics emerge and the activity presents other traits, which might be related to bistable regimes[23], however, this examination deserves future research using the presented analytical framework.

## Discussion

We have presented a quantitative analytical approach to analyze the dynamics of excitatory neural networks where the nodal rules imitate adaptive response probability with a finite memory. Results indicate that cooperative network oscillations stem from such nodes even without any inhibition. The generalization of the results to more complex topologies[24] as well as for networks with a distribution of $f_c$ among neurons or inhibition is straightforward. In addition, the analytical approach can be improved by taking into account the correlation between the activity of a node and its presynaptic nodes, however, for many cases the presented mean-field equations accurately pinpoint the oscillation frequencies. Moreover, the equations can be generalized to a continues-time version which can take into account a distribution for the delays by transforming the sums into integrals. The continues-time version of the equations is numerically and analytically more challenging, and deserves future research.

Throughout the manuscript we assumed that the links are supra-threshold, this assumption allowed the presentation of a simpler version of the equations. The generalization to the case of sub-threshold connections is possible and requires additional nonlinear equation, similar to Eq. 6, to take into account correlations between stimulations. Oscillations in neural networks with sub-threshold interactions were experimentally observed[25], and the generalization of the analytical framework to the case of delay-coupled networks with sub-threshold interactions[26] would vastly expand the horizon and potential impact of the presented framework.

Our analytical approach simplifies a problem from a complex nonlinear set of N equations to a finite small set of equations, where only the magnitude of the noise is dependent on N. For the specific case of a neural network where the activity depends on the history, in a nonlinear (multiplicative) way, Eq. (6), the resulting set of equations (Eqs. 4-8) are difficult to solve analytically and we used numerical analysis. It might be possible to solve some approximated version of the equations and compute the approximated power spectrum analytically following standard techniques[27-29].

The presented analytical approach is applicable also to other types of graphs, including scale-free networks[30]. It is possible to numerally solve such models when the graphs consist of a finite number of classes of nodes, similar to the four classes presented in Figs. 3 and 4, or sub-sampling a scale-free distribution. Such solutions might reveal an interesting interplay between the oscillation modes and the topology of the network. However, the extension of our results to an infinite number of classes (the entire distribution of the scale-free connectivity), presents a challenge which certainly deserves further research.

The examined model is based on two features, hyperactive nodes degenerate their neighbors and nodes having memory. These features are common in a diversity of real-world complex networks, where competition exists, e.g. economics[12]. Nevertheless, the exemplified framework can describe the dynamical properties of the different type of network models, by modifying Eqs. (3) and (5) according to the response of the nodes and the details of the graph, respectively. We anticipate the presented analytical framework and our results to bring a better understanding as well as an advanced methodology for the investigation of coexisting cooperative modes of activity in a variety of complex networks of networks and scale-free networks.

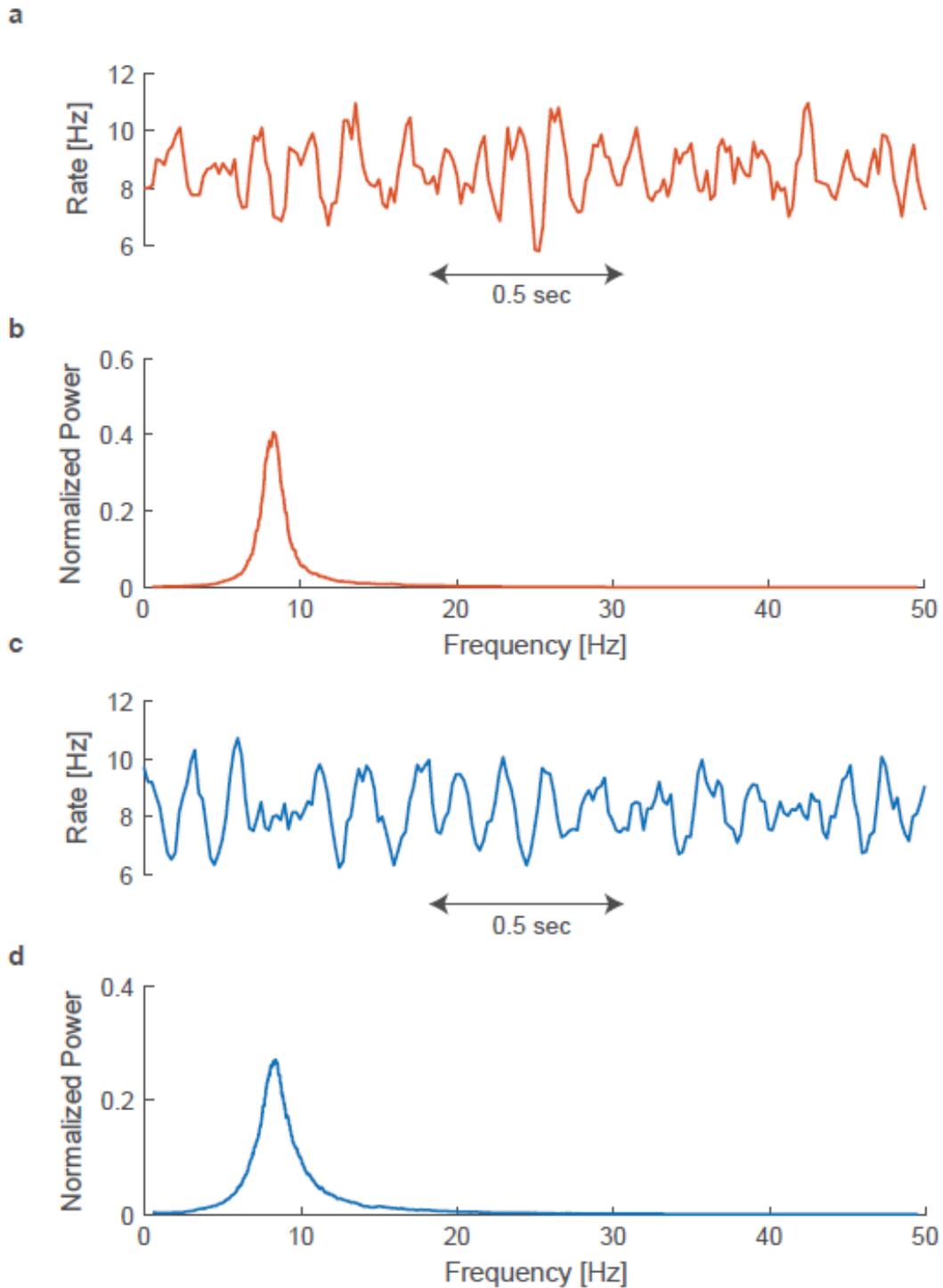

**Figure 1. An example of the dynamics of a sparse excitatory neural network with neuronal response failures.** (**a**) Simulation of Eqs. (1)-(3), exemplified by the average firing over windows of *d (=10 ms)*. (**b**) The normalized smoothed power spectrum (convoluted with a *rect(f/1Hz)*) of 200 seconds of the time-dependent average firing rate. (**c-d**) Same as (a-b) but a numerical solution of Eqs. (4)-(8), rather than a simulation.

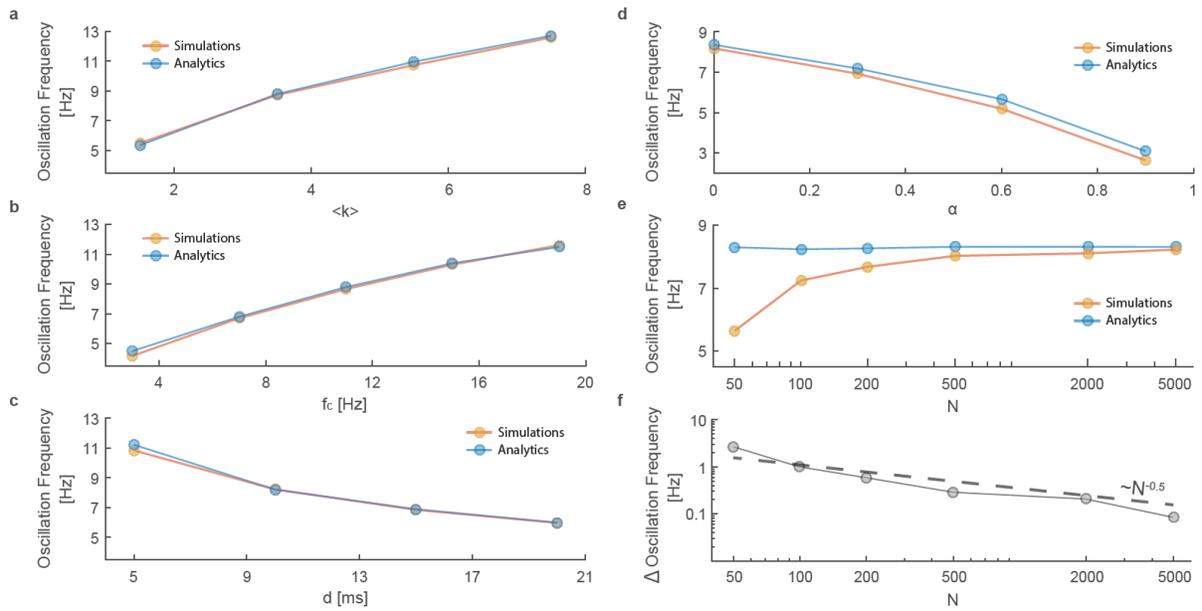

**Figure 2. The oscillation frequency where only one parameter of the network is tuned (x axis).** The oscillation frequency (**a**) increases with *<k>*, (**b**) increases with *$f_c$*, (**c**) decreases with *d* and (**d**) decreases with *α*. (**e**-**f**) Decreasing the size of the network results in finite size effects, which decreases as $N^{-0.5}$, the quantity presented in (**f**) is the difference between the two curves from (**e**). Each data point is the average of 10 different trials where the standard deviations are smaller than the circles. The oscillation frequency was defined as the maxima of the smoothed power spectrum (convoluted with a *rect(f/1Hz)* function) per trial, with a duration of 210 seconds, where the first 10 seconds were neglected (transient).

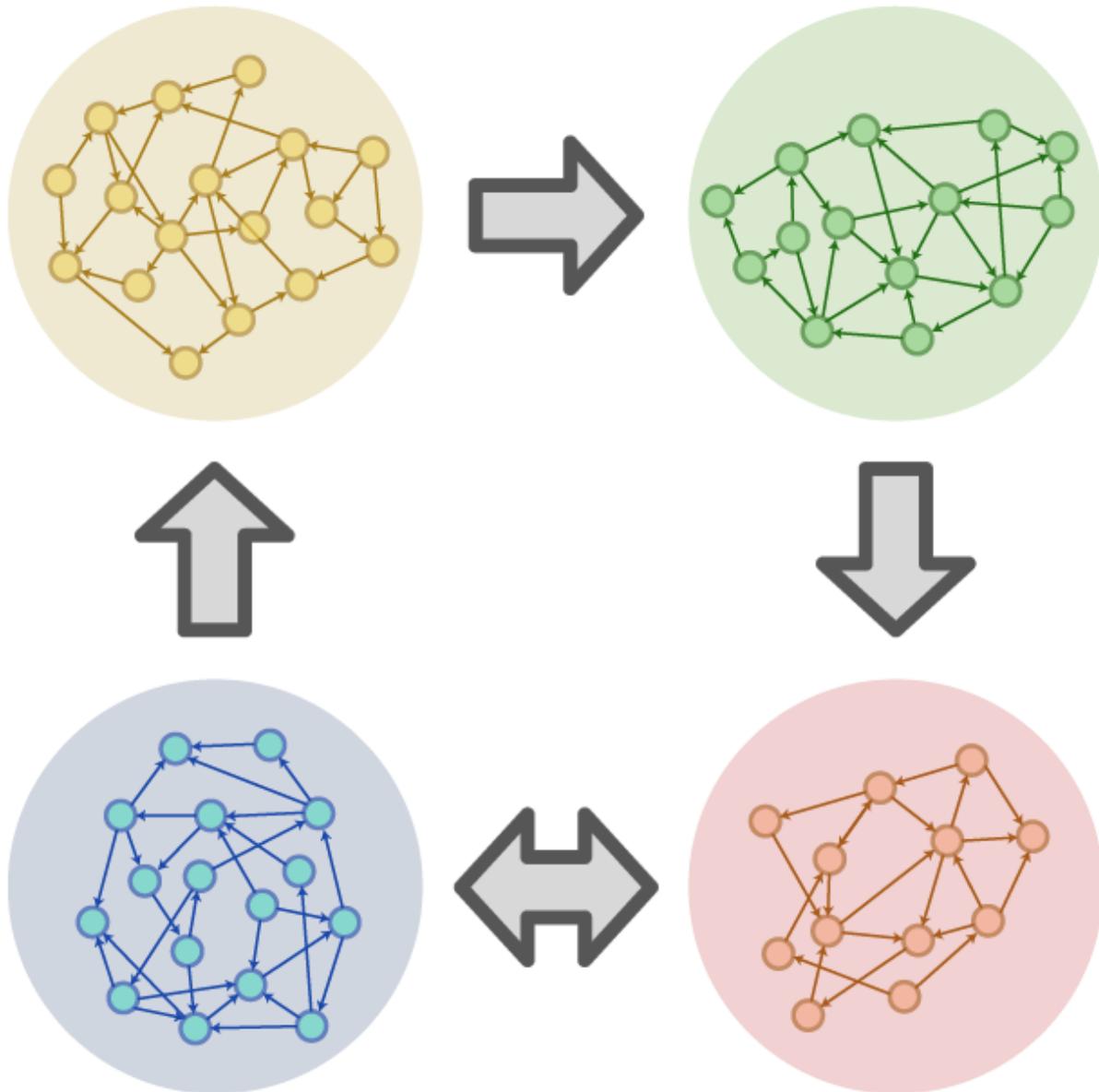

**Figure 3. Network of networks.** An illustration of a complex network of networks, composed of four sub-networks, orange, green, blue and red, which can be studied using the generalized analytical approach.

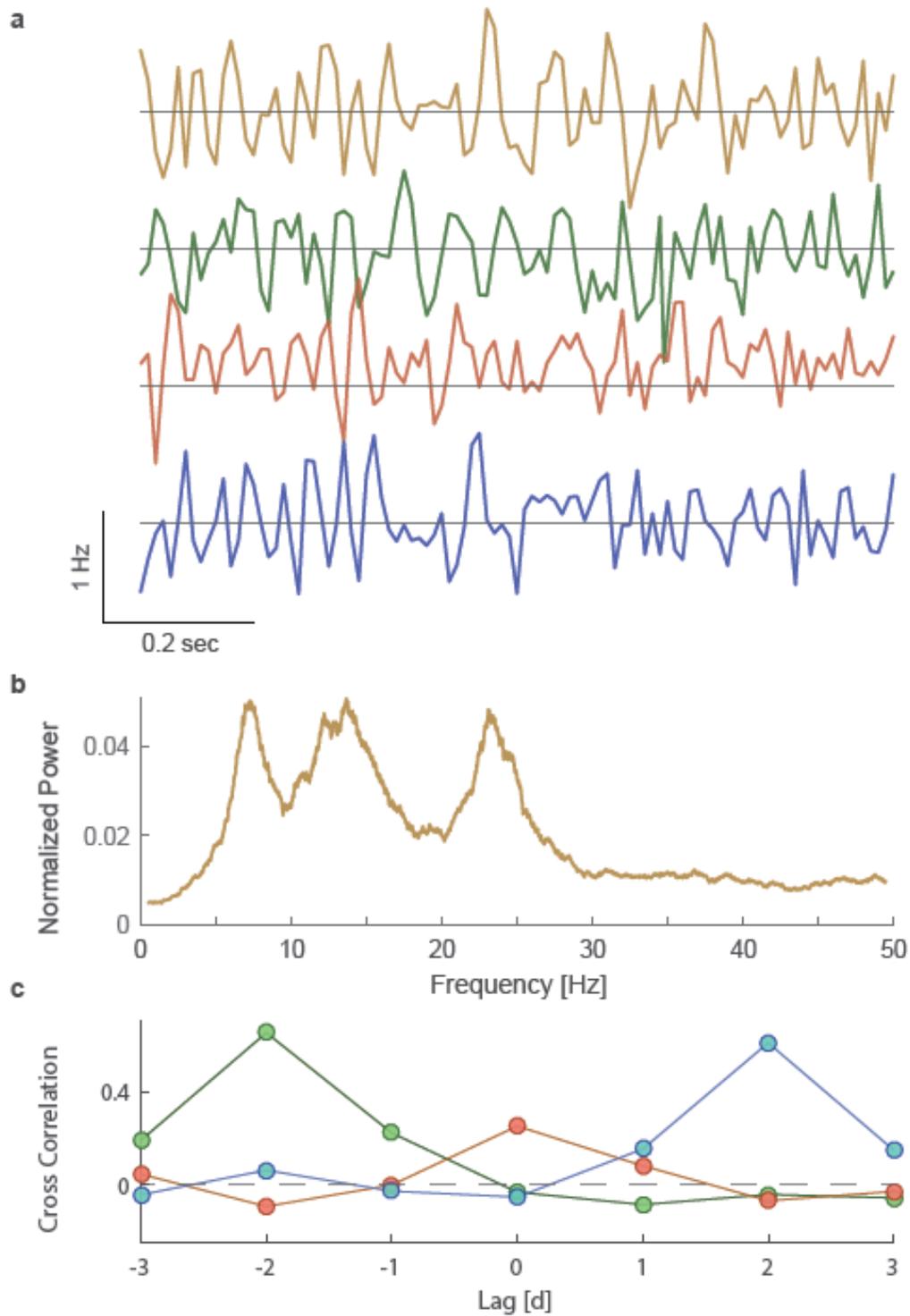

**Figure 4. Analytical analysis of a network of networks, with 2000 neurons in each sub-network, color coded according to Fig. 3.** (**a**) A typical sample of the activity, *R(i)*, of the four sub-networks. (**b**) The smoothed power spectrum (convoluted with *rect(f/1Hz)*) of the orange sub-network activity, over 210 seconds (first 10 seconds are neglected), illustrating peaks around 8, 12.5 and 25 Hz. (**c**) The cross-correlations between the activities of the orange sub-network and the other sub-networks (color coded) over 200 seconds as in (b).

-----------

**Acknowledgments**

This work was supported by the TELEM grant of the planning and budgeting Committee of the Council of Higher Education, 2015, Israel.

**Author contributions**

A.G. developed the analytical framework together with H.U. and performed the simulations. S.S performed some of the simulations. I.K. initiated the study and supervised all aspects of the work. All authors discussed the results and commented on the manuscript.

# Supplementary: Oscillations in networks of networks stem from adaptive nodes with memory

Amir Goldental[†], Herut Uzan[†], Shira Sardi and Ido Kanter[*]

**Normalization of $h_k(i,m)$**

The $h_k(i,m)$ that appears in Eq. (8) is normalized,

$$h_k(i,m) = h'_k(i,m) / \left[ q_k(i) + \sum_{m=1}^{(f_c d)-1} h'_k(i,m) \right] \quad (S1)$$

where $h'_k(i,m)$ is the $h_k(i,m)$ presented in Eq. (6) and

$$q_k(i) = 1 - \sum_{m=1}^{(f_c d)-1} \langle h'_k(i,m) \rangle + \xi_k^q(i) \quad (S2)$$

where the last term stands for random fluctuations in finite networks which scale with $N^{-0.5}$ (Gaussian random variable with a variance of $\langle q_k(i) \rangle \cdot (1-\langle q_k(i) \rangle)/(p^{st}_k(i) \cdot C_k \cdot N)$ and a zero mean).

**α>0**

In case $\alpha > 0$ the response probability is given by Eq. (3), which can be rewritten as

$$p_\alpha^{sp}(\Delta t_2, \ldots, \Delta t_n) = \alpha \cdot p_\alpha^{sp}(\Delta t_2, \ldots, \Delta t_{n-1}) + (1-\alpha) \cdot p^{sp}(\Delta t_n) \quad (S3)$$

The first term in the right hand side can be approximated, by averaging, as

$$\alpha \cdot p_\alpha^{sp}(\Delta t_2, \ldots, \Delta t_{n-1}) \approx \alpha \cdot p_\alpha^{sp}\left(\frac{d}{k\langle R\rangle_i}, \ldots, \frac{d}{k\langle R\rangle_i}\right) = \alpha p^{sp}\left(\frac{d}{k\langle R\rangle_i}\right) \quad (S4)$$

where $\langle R \rangle_i$ is given by solving Eqs. (4)-(8) while setting $R(i)=\langle R \rangle_i$ for all $i$ and neglecting all the noise terms.

**Network of networks**

Here each sub-network, $g$, has a connectivity probability mass function $C_{g,k1\ldots kn}$ which gives the probability for a neuron from sub-network $g$ to receive stimulations from $k_1,\ldots,k_n$ neurons from sub network $1,\ldots,n$, respectively. For

example, the $C_{g,k1...kn}$ which describes the system in Fig. 4 in the manuscript is given by

$$C_{g,k_1,...,k_n} = \delta_{g,1}\delta_{k_1,1}\delta_{k_2,0}\delta_{k_3,0}\delta_{k_4,2} + \delta_{g,2}\delta_{k_1,2}\delta_{k_2,1}\delta_{k_3,0}\delta_{k_4,0}$$
$$+\delta_{g,3}\delta_{k_1,0}\delta_{k_2,2}\delta_{k_3,1}\delta_{k_4,2} + \delta_{g,4}\delta_{k_1,0}\delta_{k_2,0}\delta_{k_3,2}\delta_{k_4,1} \text{ (S5.1)}$$

Additionally, the delays between network $g$ and $g'$ are now given by the matrix $d_{g,g'}$, e.g. in Fig. 4

$$d_{g,g'} = [10\ ms]\delta_{g,g'} + [20\ ms](1 - \delta_{g,g'}) \text{ (S5.2)}$$

All nodes that share the same connectivity properties ($g,k_1,...,k_n$) have the same stimulation probability $p^{st}_{g,k1,...,kg}$ and susceptibility $\chi_{g,k1,...,kg}$. As a result, the $R$ of group $g$, denoted as $R_g$, is given by:

$$R_g(i) = \sum_{k_1=0}^{k_{max}} ... \sum_{k_n=0}^{k_{max}} C_{g,k_1,...,k_n} \cdot p^{st}_{g,k_1,...,k_n}(i) \cdot \chi_{g,k_1,...,k_n}(i) \text{ (S5.3)}$$

where

$$p^{st}_{g,k_1,...,k_n}(i) = 1 - \prod_{g'=1}^{n}[1 - R_g(i - d_{g,g'})]^{k_g} \cdot (1 - f_{ext} \cdot d) + \xi^{st}_{k_1,...,k_n}(i)$$
(S5.4)

and the susceptibility is given by

$$\chi_{g,k_1,...,k_n}(i) = 1 - \sum_{m=1}^{(f_c d)^{-1}}(1 - m \cdot d \cdot f_c)h_{g,k_1,...,k_n}(i,m) \text{ (S5.5)}$$

where

$$h_{g,k_1,...,k_n}(i,m) = p^{st}_{g,k_1,...,k_n}(i-m)\prod_{n=1}^{m-1}[1 - p^{st}_{g,k_1,...,k_n}(i-n)] +$$
$$\xi^{h}_{g,k_1,...,k_n}(i,m) \text{(S5.6)}$$

Similar to the manuscript, $h_{g,k1,...,kn}$ is normalized, the variance of the stochastic term in Eq. S5.4 is
<$p^{st}_{g,k1,...,kg}(i)$>·(1-<$p^{st}_{g,\ k1,...,kg}(i)$>)/($C_{g,k1,...,kg}$·N) and the variance of the stochastic term in Eq. S5.6 is
<$h_{g,k1,...,kg}(i,m)$>·(1-<$h_{g,k1,...,kg}(i,m)$>)/($p^{st}_{g,k1,...,kg}(i)$·$C_{g,k1,...,kg}$·N)